\documentclass[%
 aip,
rsi,
 amsmath,amssymb,
 reprint,%
]{revtex4-1}

\usepackage{graphicx}% Include figure files
\usepackage{subcaption}
\usepackage{dcolumn}% Align table columns on decimal point
\usepackage{bm}% bold math
\usepackage{adjustbox}
\usepackage[percent]{overpic}
\usepackage[utf8]{inputenc}
\usepackage[T1]{fontenc}
\usepackage{mathptmx}
\usepackage{etoolbox}
\usepackage{caption}
\usepackage{xcolor}
\usepackage{comment}
\makeatletter
\def\@email#1#2{%
 \endgroup
 \patchcmd{\titleblock@produce}
  {\frontmatter@RRAPformat}
  {\frontmatter@RRAPformat{\produce@RRAP{*#1\href{mailto:#2}{#2}}}\frontmatter@RRAPformat}
  {}{}
}%
\makeatother
\begin{document}

\preprint{AIP/123-QED}

\title[Sample title]{Localized In-vacuum dry loading of ~0.1 - 10$\mu$m spherical and plate-like particles into optical traps using a pulled glass capillary \\
Andrew Dana title idea: }
\title{Localized efficient in-vacuum loading of $\sim$0.1-10 $\mu$m spherical and plate-like particles into optical traps using a pulled glass capillary}
%  \title{SGE:modified title idea? A piezo driven, pulled glass capillary-based device for localized and  efficient in-vacuum loading of nano-micron scale particles into optical traps}
% Force line breaks with \\
\author{Alexey Grinin}
 \altaffiliation{Also at Levisens, LLC, Chicago, IL.}%Lines break automatically or can be forced with \\
\author{Andrew Dana}%
 
\affiliation{Center for Fundamental Physics, Department of Physics and Astronomy, Northwestern University, Evanston, Illinois 60208, USA%\\This line break forced with \textbackslash\textbackslash
}%
\author{Mark Nguyen}%
 
\affiliation{Center for Fundamental Physics, Department of Physics and Astronomy, Northwestern University, Evanston, Illinois 60208, USA%\\This line break forced with \textbackslash\textbackslash
}%

\author{Scott Grudichak}%
 
\affiliation{Center for Fundamental Physics, Department of Physics and Astronomy, Northwestern University, Evanston, Illinois 60208, USA%\\This line break forced with \textbackslash\textbackslash
}%

\author{Katarina Boskovic Guy}%
 
\affiliation{Center for Fundamental Physics, Department of Physics and Astronomy, Northwestern University, Evanston, Illinois 60208, USA%\\This line break forced with \textbackslash\textbackslash
}%

\author{Shelby Klomp}%

\affiliation{Center for Fundamental Physics, Department of Physics and Astronomy, Northwestern University, Evanston, Illinois 60208, USA%\\This line break forced with \textbackslash\textbackslash
}%
\author{Shafaq Gulzar Elahi}%
 
\affiliation{Center for Fundamental Physics, Department of Physics and Astronomy, Northwestern University, Evanston, Illinois 60208, USA%\\This line break forced with \textbackslash\textbackslash
}%
\affiliation{Center for Interdisciplinary Exploration and Research in Astrophysics, Department of Physics and Astronomy, Northwestern University, Evanston, Illinois 60208, USA}
\author{Sam Borden}%

\affiliation{Center for Fundamental Physics, Department of Physics and Astronomy, Northwestern University, Evanston, Illinois 60208, USA%\\This line break forced with \textbackslash\textbackslash
}%

\author{Zhiyuan Wang}%
 
\affiliation{Center for Fundamental Physics, Department of Physics and Astronomy, Northwestern University, Evanston, Illinois 60208, USA%\\This line break forced with \textbackslash\textbackslash
}%

\author{George Winstone}%
 
\affiliation{Center for Fundamental Physics, Department of Physics and Astronomy, Northwestern University, Evanston, Illinois 60208, USA%\\This line break forced with \textbackslash\textbackslash
}%
\affiliation{Center for Interdisciplinary Exploration and Research in Astrophysics, Department of Physics and Astronomy, Northwestern University, Evanston, Illinois 60208, USA}
\author{Andrew A. Geraci}
 \email{andrew.geraci@northwestern.edu}%\homepage{http://www.Second.institution.edu/~Charlie.Author.}
%\affiliation{%
%Second institution and/or address%\\This line break forced% with \\
%}%
\affiliation{Center for Fundamental Physics, Department of Physics and Astronomy, Northwestern University, Evanston, Illinois 60208, USA}
\affiliation{Center for Interdisciplinary Exploration and Research in Astrophysics, Department of Physics and Astronomy, Northwestern University, Evanston, Illinois 60208, USA}

\date{\today}% It is always \today, today,

\begin{abstract}

We demonstrate a compact piezoelectric-driven micropipette launcher for localized in-vacuum delivery of nano- and microparticles into optical traps. The launcher has been integrated into multiple optical trapping setups, including a single-beam trap, a non-interfering dual beam trap, and a standing-wave dual beam trap, showcasing the versatility and ease of integration of the setup. Using the micropipette launcher, we have successfully trapped silica spheres of $170\text{ nm}$, $300\text{ nm}$, 3 $\mu\text{m}$ diameter, as well as 6 $\mu\text{m}\times$ 0.2 $\mu\text{m}$ $\beta$-NaYF hexagonal prisms and $\sim 100$ nm diameter high-purity nanodiamonds. We characterize the performance of the device including the peak acceleration, angular distribution of emitted particles, and the dependence on vertical displacement between the pipette tip and optical trap. Trapping efficiency as high as 93\% is achieved. %A typical micropipette loading of some of these particles leads to hundreds of successful trapping events with as high as 93\% efficiency. %Even a filling with as little as one hundred thousand $\beta$-NaYF hexagonal prisms in the pipette still leads to multiple successful trapping events.

%We describe a piezoelectric driven apparatus for in-vacuum loading of nano- and micro-particles into an optical trap in vacuum using a pulled glass capillary. The device provides localized particle delivery and facilitates efficient loading of the trap from mm-range distances. We demonstrate loading of silica spheres of diameters 170 nm, 300 nm, and 3 $\mu$m as well as $\beta$-NaYF hexagonal prisms and nanodiamonds into a variety of optical traps at $\sim$ mbar pressures including single beam tweezers and dual beam configurations. 
\end{abstract}

 \maketitle

\section{\label{sec:level1}Introduction}

Advances in the field of atomic, molecular, and optical physics have benefited from having a rich history of trapping particles from the atomic to the micron size scale. Exploiting various techniques like optical trapping, ion trapping, and magnetic trapping, scientists have been able to study various physical phenomena with an extremely high level of precision. This has led to the development of cold-atom based sensors such as atomic interferometers \cite{atomintfreview} and ultra-precise clocks\cite{clocksreview}, as well as the exploration of a multitude of condensed matter physics phenomena involving Bose Einstein condensates \cite{becreview} and Fermi degenerate gases \cite{fermigasreview}. The field of levitated optomechanics specifically, where trapping of nanometer to micrometer sized particles in vacuum environments is possible \cite{Millen_2020,reviewmarkus,Tongcang_review_2013}, has shown great promise for studying fundamental physics at the table-top scale \cite{moore2021searching}. Such levitated systems have been cooled into the quantum regime \cite{delic2020cooling,Marin2022,tebbenjohanns2021quantum,Novotny2022, groundstate}, enabling the control of quantum behavior at macroscopic scales and potential applications in quantum information science and quantum sensing  \cite{Millen_2020,reviewmarkus}.  

%Before one can study interesting physics with these levitated systems, one must consider
An important aspect of realizing these levitated systems is the preparation and loading of nano- and micro-particles into the trap. Current state-of-the-art particle loading techniques include the use of nebulizers \cite{AspelmeyerCooling,Novotny2012}, piezoelectric transducers \cite{weisman2022,khodaee2022,Li:2011}, laser-induced acoustic desorption (LIAD) \cite{MillenLIAD,northup}, transferring a particle to ultra-high vacuum in a hollow-core fiber  \cite{HCFload}, or particle release by sublimation of camphor \cite{dursocamphor}. 
Several of the advantages and limitations of these methods have recently been summarized in Ref. \cite{ronpaper}. The applicability of different techniques in some cases depends on the specific types of trap (e.g. ion trapping, magnetic trapping, or optical trapping), considering attributes such as typical trapping depth, trapping volume, and operating environmental pressure and temperature conditions.  

In certain cases it is favorable to implement a loading mechanism that is vacuum-compatible --- such as using a piezoelectric transducer, hollow-core fiber transport, or light-assisted desorption --- to eliminate the need to repump the vacuum chamber between experimental runs. This is advantageous for certain applications, including measurements involving repeatedly dropping particles from a trap, matter wave interferometry \cite{Ulbricht:2014,GeraciMatterwave,kaltenbaek2016macroscopic,Kaltenbaek_2023}, as well as measurements involving loss of the particle from the trap, e.g. to study heating mechanisms or for trap stability studies. In the direction of constructing a matter wave interferometer with nanometer scale particles, the necessity for an approach that can reliably and controllably load thousands of particles while also being cryogenic, ultrahigh vacuum compatible naturally arises. 

Previously, our research group \cite{weisman2022} and others \cite{khodaee2022,Li:2011,ronpaper} have developed methods involving shaking particles from a glass substrate consisting of a commercially available microscope slide \cite{weisman2022} or microscope cover slip \cite{Li:2011} that is positioned above the trapping region by use of a hard ceramic PZT ring. Unlike the LIAD based approach, which has been used to successfully load nanoparticles into ion traps at high vacuum, the piezo-based method allows for comparably lower center-of-mass velocities of the particles as they reach the trapping region, which is beneficial for not exceeding the capture velocity of an optical trap.  
In this paper, we describe a new piezoelectric transducer based method to enhance the in-vacuum loading efficiency of nanometer- to micrometer-sized particles % with varying materials and of varying geometries 
into optical traps in a more localized fashion by use of a pulled glass capillary tube.

By utilizing the relatively sharp tip of the micropipette which has a diameter of approximately $10-30$~$\mu$m, it is possible to position the particles in closer proximity to the optical trap than is possible for flat plate based piezo-based methods, due to the often significant divergence angle of tightly focused beams used in optical trapping. In cases where terminal velocity is not achieved, often at lower pressures or for larger diameter particles, this typically results in smaller particle velocities due to the decreased falling distance under the acceleration due to the Earth's gravity. 
%\textcolor{black}{assuming terminal velocity is not achieved}. 
Furthermore, by extending the small diameter ($<1$~mm) capillary tube into a cryogenic vacuum region, it is possible to efficiently deliver particles close to an optical trap contained within a cryogenic volume, with the piezoelectric transducer operating outside of the radiation shields at room temperature.  

Piezo-based loading methods are typically limited by the need to overcome stiction forces, which become challenging to overcome for particle sizes below $100$~nm. %\textcolor{teal}{be more quantitative on efficiency}
We find that the pulled glass capillary-tube based approach is more efficient than flat plate substrate methods previously used in our lab, and is able to launch particles of comparably small diameter to what has been done with the flat plate based substrate used in earlier works. In particular we report trapping efficiencies as high as $93$\% per trial of the piezo-actuator for loading individual or aggregates of $300$~nm silica spheres with the tip positioned approximately $4$~mm above a dual beam trap at mbar pressures, approximately an order of magnitude more efficient than previously used methods involving flat plate substrates at $\sim 2-3$ cm-distances \cite{weisman2022}. Furthermore we study the angular distribution of emitted particles and find that the flux is localized to a cone of less than a $10$ degree opening angle.
We additionally find that for smaller pipette opening diameters approaching $10$~$\mu$m, the pipette tip tends to become obstructed by particles in the size range that we explored, whereas a $30$~$\mu$m diameter tip works robustly for all of the particle types and sizes attempted in our experiments.

%Although the results presented in this paper involve optical traps, it is not necessary to limit its application to just optical trapping since in general, piezoelectric transducer methods have proven to be successful e.g. in Paul traps.

This paper is organized as follows.  In Section \ref{setup} we describe the construction and operation of the piezo-based micropipette particle loading mechanism.  In Section \ref{Sec:results} we describe the results of operating the device and analyze its performance and efficiency to load optical traps in single beam tweezer and dual beam counter-propagating configurations for different pipette tip sizes and for spherical and nonspherical dielectric particles with sizes ranging from 100~nm to several microns and of differing material composition.  We conclude with an outlook for possible uses of the device in other types of particle traps and particle delivery systems.

 %This recent paper shows the successful loading and trapping of nanodiamonds using a piezoelectric transducer albeit of ring geometry compared to the cylindrical geometry contained in this new method.[insert folman launching paper reference]. Existing experimental setups in our lab previously utilizing this ring piezo method have experienced a significant increase in loading and trapping rates upon installation of this new device.
%\textcolor{red}{AG: add something about stiction forces etc..}

\section{Experimental Apparatus \label{setup}}
\subsection{Device Design and Electromechanical Characterization}
\begin{figure}[h]
    \centering
    \includegraphics[width=0.45\textwidth]{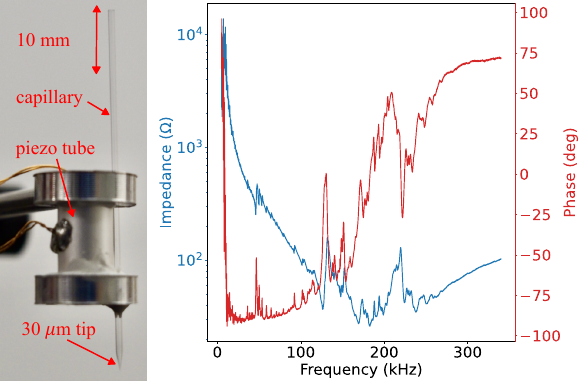}
    \caption{\textbf{Left:} Micropipette launcher consisting of a piezo tube (Thorlabs PT49LM, $10~\text{mm}$ length, 2.8~$\mu \text{m}$ axial free stroke, $14~\text{nF}$ capacitance), electrical wires, upper aluminum base plate with a through hole for the pipette, lower aluminum plate to mount (glue) the micropipette and the micropipette (WPI, model TIP30TW1, 30~$\mu\text{m}$ opening). \textbf{Right:} Amplitude and phase spectrum of the electrical impedance of the micropipette launcher, showing several impedance minima. A global impedance minimum of approximately 26~$\Omega$ is observed at 182.9~kHz.}
    \label{fig:CAD}
\end{figure}

The piezoelectric transducer-based particle disperser uses a micropipette with a small tapered opening to achieve a directed, localized, and controllable stream of particles, while providing a large reservoir of particles in the main volume of the pipette. In this way, very small amounts of particles or alternatively very many launches without refilling can be achieved, crucial to many scientific and engineering applications. By minimizing the mass of the piezo load, we allow for high-frequency oscillation to maximize the acceleration. When driven sinusoidally at  a frequency $f$, the peak acceleration $a$ scales quadratically with the oscillation frequency as
\begin{equation}
a =  (2\pi f)^2 x,
\label{acceleration_eqn}
\end{equation}
where $x$ is the (frequency-dependent) displacement amplitude of the oscillation. For efficient trapping, we typically operate close to the mechanical resonance $f_0$ of the piezoelectric transducer (i.e., $f\simeq f_0$) to maximize $x$ at a given drive voltage. In order to provide high acceleration to overcome stiction forces of nanometer-sized particles to adjacent surfaces, high resonance frequencies are beneficial. Applying a load of mass $M$ to a piezoelectric transducer of mass $m$ and eigenfrequency $f_0$ generally reduces the resonance frequency to $f_{0}'$; in a simple effective mass-on-spring model (distributed actuator mass approximation \cite{fox1970effective})
\begin{equation}
    f_{0}' =  f_{0}\sqrt{\frac{m/3}{m/3+M}}.
\end{equation}
Thus the mass of the pipette and the mounting block (see Fig.~\ref{fig:CAD}) needs to be minimized. High oscillation frequencies also necessitate low effective capacitance $C$ to avoid limiting the high-frequency drive amplitude (voltage $V$ and current $I$) of the high-voltage amplifier which drives the piezoelectric transducer ($I\propto 2\pi f C V$ for a capacitive load). On the other hand, the amplitude (free stroke) of a piezoelectric actuator increases with length. Piezoelectric tube actuators provide a good compromise as they combine relatively large free strokes with high nominal resonance frequencies and low capacitances. 

Figure \ref{fig:CAD} shows the assembly of the pipette launcher consisting of machined cylindrical aluminum upper and lower pieces, a piezoelectric transducer tube (Thorlabs PT49LM, $10~\text{mm}$ length, $2.8~\mu \text{m}$ axial free stroke, $14~\text{nF}$ capacitance), 32~AWG copper wires soldered to the inner (+) and outer (-) sides of the piezo tube and a pulled glass capillary tube (WPI, model TIP30TW1\cite{wpi_tip30tw1}, 30~$\mu\text{m}$ tapered opening, 1~mm OD, 0.75~mm ID). The pipette is glued to the lower aluminum piece and goes through a concentric, larger hole  (2~mm) in the upper plate without touching it to allow free vibration. Vacuum compatible epoxy (TorrSeal) is used both to glue the aluminum plates to the piezoelectric tube and the glass capillary to the lower plate. The holes in the center of the upper and lower plates (5~mm diameter) provide access to the inner (positive) electrode and reduce load mass. % and can serve as window for a microscope or other sensors. 
Optionally, in some setups, we have made $1.5$~mm grooves in the lower plate to attach a protective cover to prevent particle deposition on nearby surfaces. Once in a vacuum chamber, rough alignment of the micro-tip can be done by placing it near the trap by eye. More precise alignment is achieved with vacuum compatible manual translation stages or motorized stages for in-situ alignment. 

\subsection{Acceleration Measurement}
\begin{figure}[h]
    \centering
    \includegraphics[width=\linewidth]{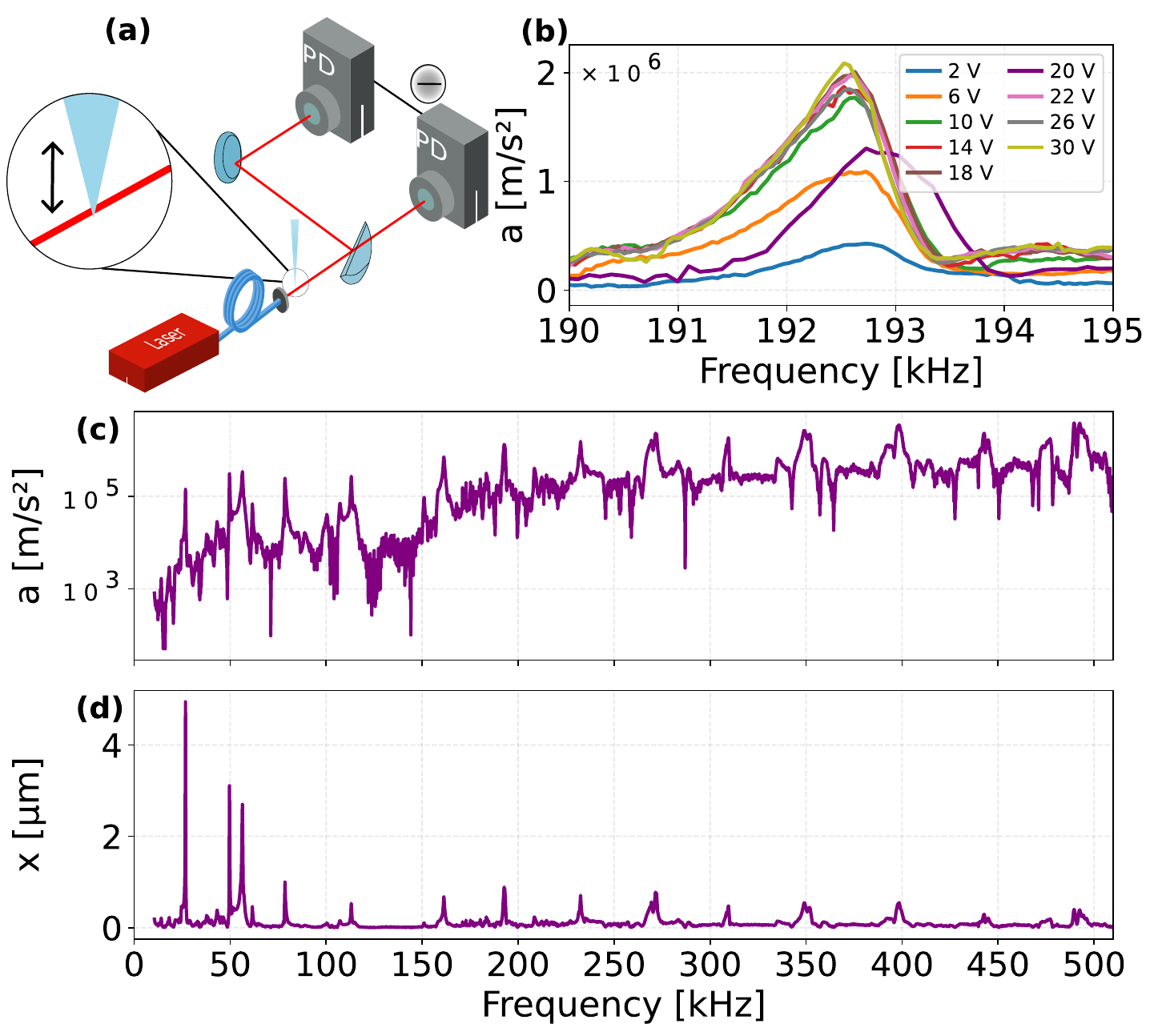}
    \caption{\textbf{(a):} Vertical pipette motion measurement setup. Acceleration was inferred by multiplying the displacement spectrum with $(2\pi f)^2$. The setup employs a D-Mirror with a balanced photodiode to detect the motion of the pipette tip cutting through a collimated Gaussian beam. \textbf{(b):} Study of acceleration as a function of the peak driving voltage near the resonance peak around $192.6$~kHz. \textbf{(c):} Acceleration versus drive frequency for a drive voltage of 20~V. \textbf{(d):} Displacement versus drive frequency for 20~V drive voltage.}
    \label{fig:displacement_accel}
\end{figure}

Figure \ref{fig:displacement_accel} shows the measured spectrum of the vertical motion (d) of the micropipette, the inferred acceleration spectrum (c), peak acceleration for different driving voltages (b) and the measurement setup (a). We placed the pipette tip into the center of a $2$~mm diameter collimated $1560$~nm beam and swept the frequency of the high-voltage piezo drive (20~Vpk, 100~V offset, A.A.Lab Systems A-301 HS) from $1$~kHz to $510$~kHz. By placing a D-mirror with the flat edge perpendicular to the direction of the pipette, and a balanced photodetector (Thorlabs PDB250C) behind the pipette, the vertical motion was detected and demodulated at each frequency using a lock-in amplifier (Zurich Instruments MFLI). A voltage-to-distance calibration was performed prior to the measurement by mounting the pipette launcher on a vertical manual translation stage. The pipette motion was recorded in vacuum ($0.7-2.5$~mbar) corresponding to the typical trapping pressure. A more confined sweep from $190$~kHz to $195$~kHz was performed to study the relationship between maximal acceleration and drive voltage, as shown in Fig. \ref{fig:displacement_accel} (b).

\subsection{Particle Preparation}
The particles tested with this loading apparatus are either commercially bought from suppliers such as Bangs Laboratories for SiO$_2$ spheres, Adamas Nanotechnologies for nanodiamonds, or provided by collaborators in the cases of doped and undoped $\beta$-NaYF hexagonal prisms \cite{NAYFtrapping} and high purity nanodiamonds \cite{MilledND}. For  SiO$_2$ spheres, nominal diameters of $170$~nm, $300$~nm, and $3$~$\mu$m have been tested. Figure \ref{fig:pipettelaunchercollage} visualizes the range of particle species and sizes demonstrated with this launching method in comparison to other methods of in-vacuum loading of nano- and microparticles into optical traps. Of the particles loaded with this micropipette launcher, two different preparation methods have been used with success, which can both be done before or after the micropipette is glued into the piezo assembly. Commercial nanoparticles often come in a colloidal suspension of deionized water or isopropanol. In these cases, where there is an abundance of particles, they are prepared by pipetting a small amount onto a glass substrate which is placed on a hotplate at $170$ $^\circ\mathrm{C}$ with a heat lamp to evaporate the liquid leaving an amorphous solid. This is then crushed into a powder either with a tungsten rod or an agate mortar and pestle. Once a fine powder is produced by iteratively breaking up the solid, the pulled glass capillary is loaded with particles from the end with the larger opening by gently pressing the end into the powder repeatedly. By occasionally turning the capillary over and using a wire smaller than the opening, the compacted powder can be broken up further allowing it to fall towards the micro-tip. This process is repeated until the desired amount of powder is in the glass capillary. 

The second preparation method, used primarily for particle species which are particularly valuable or have a small supply such as hexagons and high purity nanodiamonds, is to create a liquid solution of $100\%$ isopropanol with a desired amount of particles added. This liquid solution is then loaded via capillary action. Using a hotplate with a heat lamp, the liquid is evaporated leaving behind the small quantity of particles. A small wire or probe tip can be used to break the meniscus formed in the capillary to allow the liquid to evaporate more rapidly. This method has been tested on doped $\beta$-NaYF hexagonal prisms which are not commercially available and come in small numbers compared to commercially bought nanoparticles. By assuming uniform size of these hexagonal prisms placed in a given volume of isopropanol, we are able to place of order $10^5$ particles into a glass capillary and still demonstrate multiple successful optical trapping events. We anticipate that successful trapping with far fewer particles is possible with this launching method. In comparison, this is orders of magnitude less than typical numbers of particles using previous piezo driving \cite{weisman2022},\cite{khodaee2022} which rely on having order $10^9-10^{12}$ particles.

\section{Results \label{Sec:results}}
\begin{figure}[h]
    \centering
    \includegraphics[width=\linewidth]{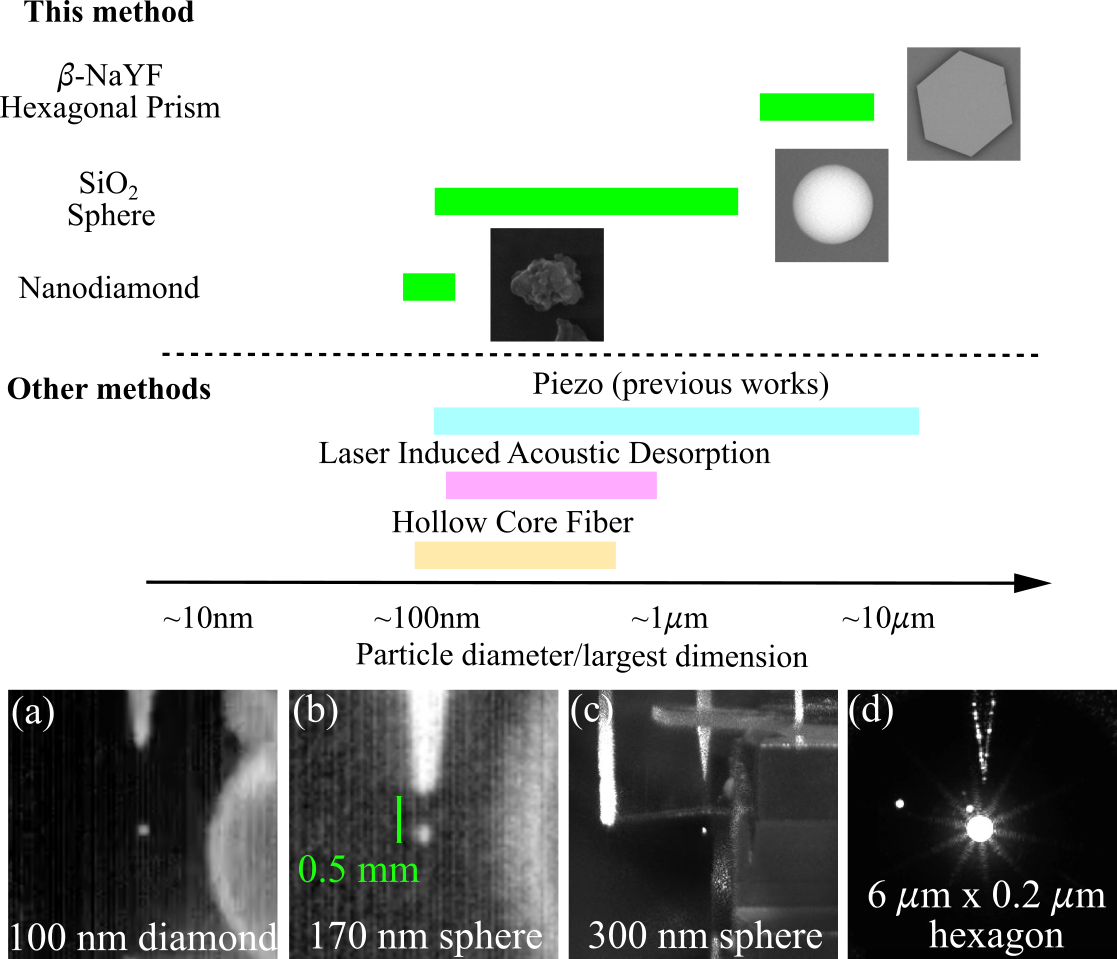}
    \caption{\textcolor{black}{\textbf{(Top)} Illustrative chart showing the size and shape of objects launched and trapped using the pulled glass capillary launcher compared to other methods for launching particles into optical traps in moderate to high vacuum. The horizontal axis refers to the approximate particle diameter or largest dimension for non-spherical shapes. For other methods, we refer to piezo based launchers \cite{weisman2022,khodaee2022,Monteiro:2020qiz} for particle diameters in the range of $100$~nm up to $20$~$\mu$m. Ref. \cite{MillenLIAD} demonstrates optical trapping of $300$~nm silica spheres and silicon rods of largest dimension $900$~nm via laser induced acoustic desorption (LIAD). Ref. \cite{HCFload} demonstrates optical trapping of silica spheres with diameters in the range of $100 - 800$~nm using a hollow core fiber. \textbf{(Bottom)} Infrared camera photographs of different size and shaped optically trapped particles loaded via the pulled glass capillary shown in the picture. Pictures \textbf{a-c} show high purity nanodiamonds and silica spheres with diameters of $100$~nm, $170$~nm, and $300$~nm respectively. Picture \textbf{d} shows a hexagon with $6$~$\mu$m diameter and $0.2$~$\mu$m thickness trapped in a counter-propagating standing wave optical trap. In \textbf{(b)}, the tip is placed $0.5$~mm away from the trapping location and still traps successfully. Details of the experimental conditions can be found in Sec. \ref{Sec:results}.}}
    \label{fig:pipettelaunchercollage}
\end{figure}

We describe the results of trapping a variety of particles using this micropipette launching method. The sizes and shapes are summarized in Fig. \ref{fig:pipettelaunchercollage} and this method is compared to other methods of loading into optical traps in Table \ref{tab:table3}.  In the following subsections, we describe the trapping efficiency, angular distribution and vertical capture profile for different heights of the pipette above the trap. 
%\textcolor{red}{ Scott: rewording trapping efficiency vs height.}

\begin{table*}[t]
\caption{\label{tab:table3}A collection of the existing dry and aerosol methods for loading nano- and microparticles into optical traps. A detailed table including loading into others kinds of traps can be found in Ref. \cite{ronpaper}. Displayed is the metric used in the literature to define the trapping efficiency and the stated efficiency as well as the pressure at which the method has been demonstrated and is typically used. The final column addresses the perceived primary limitation of each method.}
\begin{ruledtabular}
\begin{tabular}{ccccc}
 Loading Method& Efficiency Metric & Efficiency & Pressure demonstrated & Primary Limitation\\ \hline
 Piezo (this work)& Trapping event per launch & 93\% & $\sim 1$ mbar & Requires delicate handling of pipette micro-tip\\
 Piezo (previous work)\cite{weisman2022,khodaee2022}& "Sufficient Flux" 
 & Not in literature & $\sim 1$ mbar &Requires large numbers of particles\\
 LIAD\cite{MillenLIAD}& Trapping event per launch & >90\% & $\sim 1$ mbar & Requires pulsed laser, heats substrate\\
 Nebulizer (aerosol)\cite{MillenLIAD}& Loading rate & $\sim$ 1 minute & 1 bar & Wet aerosol not vacuum compatible\\
 Hollow core fiber\cite{HCFload} & Transfer efficiency& >99\% &$~10^{-9}$ mbar & Requires additional trapping prior to transfer\\
\end{tabular}
\end{ruledtabular}
\end{table*}

\subsection{Trapping Efficiency}

\begin{figure}[h]
    \centering
    \includegraphics[width=\linewidth]{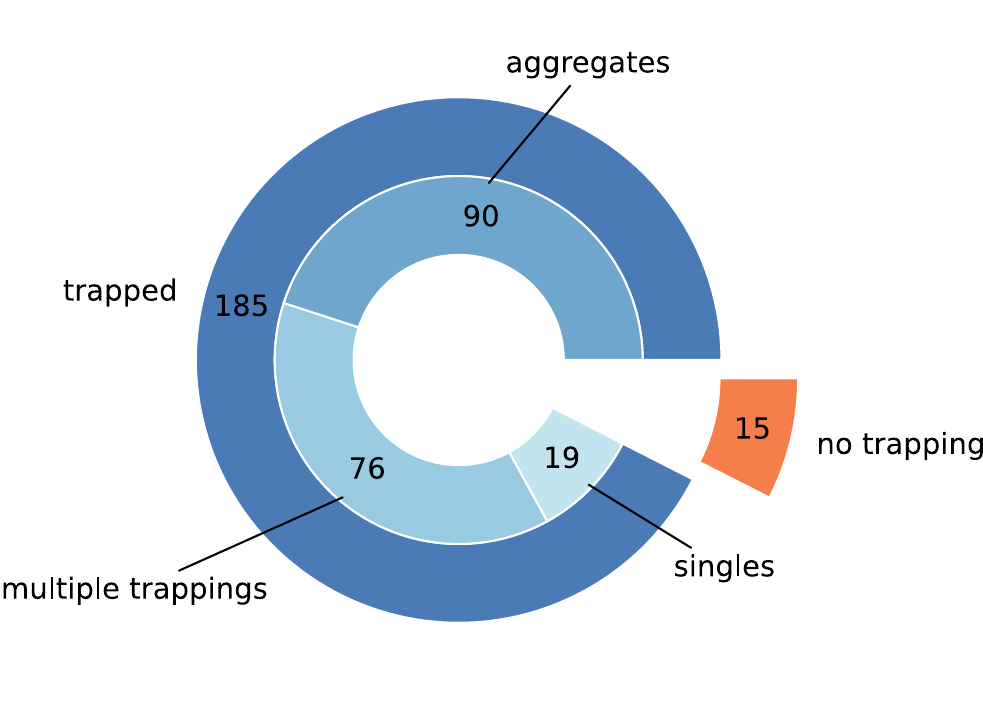}
    \caption{Trapping efficiency for $300$~nm SiO$_2$ spheres in a counter-propagating dual beam trap. Details of the experimental conditions can be found in Sec. \ref{Sec:results}. A trapping efficiency of $93\%$ was measured, with a majority of the trapped objects consisting of aggregates of spheres. }
    \label{fig:SRGpipettelauncher}
    %\caption{\textcolor{teal}{}}
\end{figure}

To judge the performance of this device, we define trapping efficiency as the number of times something is trapped per instance of launching. This definition is consistent with previous literature, for example in Ref\cite{MillenLIAD}. The trapping efficiency in a dual counter-propagating optical trap, as described in Ref. \cite{ranjit_attonewton_2015}, for $300$~nm diameter SiO$_2$ spheres is shown in Figure \ref{fig:SRGpipettelauncher}. The foci of the counter-propagating $1064$~nm wavelength beams are offset by $100$~$\mu$m, and each beam delivers $\sim350$~mW of power with a $6.5$~$\mu$m waist radius. For each launching event, the vacuum chamber pressure is held at $1$~mbar and the driving frequency of the piezoelectric tube is swept from $5$ kHz to $240$ kHz in approximately 4 seconds with a maximum voltage of $12$~V$_{RMS}$ and maximum current of $0.5$~A. Of the $200$ launching events, $185$ trapping events occur; of the particles that are trapped, $90$ are aggregates of particles, $76$ are events where multiple particles get trapped in a single launch, and $19$ are single $300$~nm spheres. This gives a total $93$\% trapping efficiency. These results were obtained with a pipette tip diameter of $30$ $\mu$m. %We note that for smaller pipette opening diameters approaching $10$ $\mu$m, the pipette tip tends to become obstructed by particles in the size range that we explored, whereas a $30$ $\mu$m diameter tip works robustly for all of the particle types and sizes attempted in our experiments.

The large number of multiple trapping events suggests that sweeping over such a large range of frequencies with these voltage and current settings leads to an abundance of particles entering the trapping region in just one launch. We have found that sweeping over smaller frequency ranges and reducing the voltage and current settings can reduce such multiple trapping events. The large number of aggregate trapping events suggest that the preparation method of these spheres can be optimized to produce more single particles via techniques such as sonication. Additionally, the number of aggregate trapping events is a result of the fact that the dual counter-propagating optical trap can confine particles over a large size and mass range. Typical optical tweezers filter out such aggregates by only being able to trap smaller particles. Based on investigations with a driven electrode, like as described in Ref. \cite{grinin_optically_2026}, we note that $\sim 90\%$ of trapped single particles have non-zero net charge.

We were able to achieve this high trapping efficiency for 300 nm SiO$_2$ spheres by optimizing the launch settings to send many particles through the trap in a single launch attempt. In some cases, such as when we are using particles for which we don't have access to large quantities, it is preferable to use settings that will cause fewer particles to fall through the trap at once. In this case, we use lower voltages or single frequency pulses instead of sweeping over many frequencies. For this reason, we often opt for lower trapping efficiencies (i.e. fewer particles trapped per launch attempt) in these cases as a trade off for having minimal wasted material. This is made possible by the ability to precisely place the tip of the micropipette millimeter distances from the trap with a highly localized and directed spray. An example of such a case is in Section \ref{heighttests} when trapping $\beta$-NaYF hexagonal prisms.

\subsection{Angular Spread of the Falling Particles} 

An advantage of the micropipette launcher over previous methods is the small angular spread of the particles as they fall toward the trap which allows for a high percentage of the falling particles to pass through the trappable region. We are able to estimate this angular spread using the distribution of particles that land on a microscope slide positioned below the trap.
%For particles that are extremely expensive or experimentally intensive to produce, the ability to be reused (incase they don't get trapped) is quite valuable and economic. 
%In our optical trapping setups with pipette launchers, a microscope slide sits at the bottom of the chamber (across from the pipette), 
%This slide is placed to cover the bottom imaging lens so that we have the ability to clean up the buildup of particles that do not get trapped during launching attempts without damaging the lens. 
This additionally allows for the possibility of reusing particles that are particularly expensive or experimentally intensive to produce.
In a dual-beam counter-propagating standing-wave optical trapping setup, as described in Ref. \cite{NAYFtrapping} with $\beta$-NaYF hexagonal prisms of $6$ $\mu$m diameter and $200$ nm thickness, we allowed the excess launched particles to collect on a slide $\approx  52.7$ mm below the pipette for several months while regularly launching around 4 mbar pressure with the launching parameters on the piezo ranging from 1-3 $V_{RMS}$ driving amplitude and 0.5A max current at different resonance frequencies. The distribution of particles is found to be slightly elliptical, so we performed a Gaussian fit along the two axes of the ellipse and used the $2\sigma$ width of the distribution to determine
%With the micro-pipette launcher, using ImageJ software
%    (\textit{side note: I used a 1-d gaussian fit along each axis in imageJ to obtain 2 sigma radii which i then used to calculate the angles reported below})
%, we have observed in our setup with NaYF particles where the pipette sits at $\approx  57.98$ mm away from the slide, 
the angular spread of the particles to be $\approx 4.6^\circ$ and $\approx  8.8  ^\circ$ respectively in the two directions. The reason for the elliptical distribution could be that for this dataset the tip was slightly chipped and hence didn't have a perfectly circular cross-section. Alternatively, since the pipette is mounted off-axis from the center of the piezo cylinder (see Fig.\ref{fig:CAD}), upon operation there may be a radial component to the velocity of the tip, leading to a greater angular spread in that direction which is consistent with the observed particle distribution.  %due to the pipette being slightly angled with respect to the bottom of the chamber and 
%due to the tip being slightly broken. 
%This small angular spread ensures that all untrapped particles land on the microscope slide and therefore can be reused with minimal loss.

% --- START OF SCOTT'S SECTION ---

\subsection{Vertical Capture Profile}
\label{heighttests}
%\textcolor{red}{MN General comments: I would reorganize these paragraphs so they flow well. It should be motivated at least at the beginning of this section and ideally emphasized throughout this section why we care about loading at different heights. A compelling story you could tell is that we want to minimize how many of these high value "particles" get released with each launching event, but maintain or better yet maximize trapping events.}

\begin{figure}[htbp]
    \centering
    \includegraphics[width=\columnwidth]{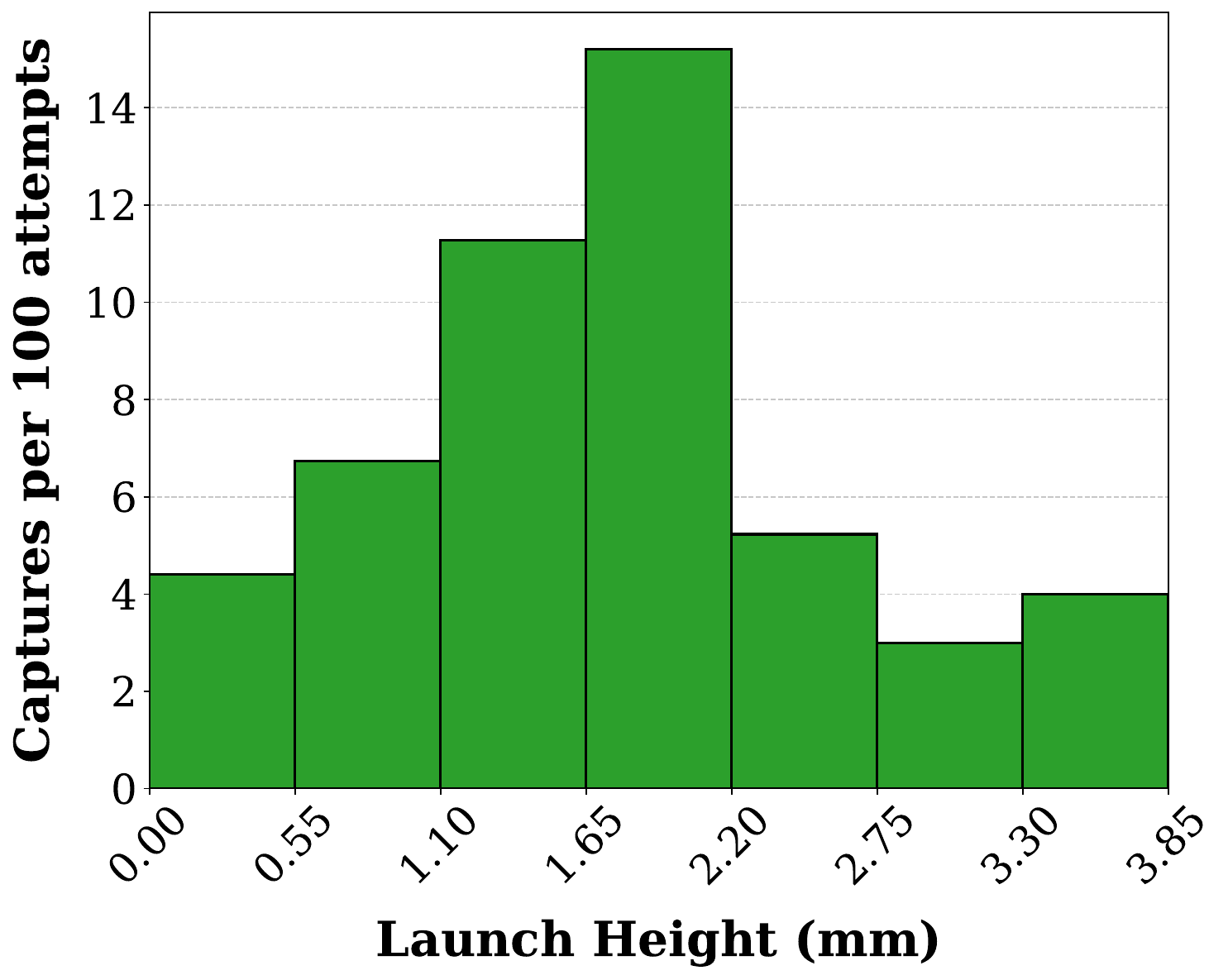}
    \caption{Vertical capture profile of $\beta$-NaYF hexagonal prisms. Launch heights are grouped into 0.55 mm bins to encompass the positional uncertainty of $\pm$ 0.27 mm. The observed capture rates indicate a broad effective operating range, with a peak near 2 mm.}
    \label{fig:prob_graph}
\end{figure}

The launch height of the pipette tip above the trapping region is a parameter that we want to optimize, especially in the case where we are using particles for which we only have access to small quantities. The goal is to determine the range of launch heights that maximize the number of capture events, even while operating at settings that do not result in the highest possible trapping efficiency. For this reason, we measure the capture profile of $\beta$-NaYF hexagonal prisms
%\textcolor{red}{Why are they called prisms here? Let's stay with a consistent naming convention so the reader isn't confused.} 
%(diameter $\approx$ 6.2 $\mu$m, thickness $\approx$ 200 nm) 
at different heights of the pipette tip above the trapping region to determine the region where the capture rate was the highest. 

% We perform this study in a dual-beam counter-propagating optical trap using a laser of wavelength $1550$ nm 

We perform this study in a dual-beam counter-propagating standing wave trap using a laser of wavelength $1560$ nm where each beam delivers $\sim150$~mW of power with a beam waist of approximately $11.5$ $\mu$m, as described in detail in Ref \cite{NAYFtrapping}.
%\textcolor{red}{\bf{[AAG: check these trap parameter numbers]}}
In order to align the pipette tip above the center of the trap and to a desired height, the pipette launcher is attached to three piezoelectric inertia stages (Newport Agilis AG-LS25) which allow for in-vacuum alignment.
%This launching technique
%\textcolor{red}{I wouldn't say setup. I would probably say something like launching technique or something similar.} 
%favors the dispersion of a small number of particles to isolate single particles. 
We used single-frequency pulses $1$--$3$~V$_{RMS}$ from the launching piezo to restrict the number of particles traversing the trap per attempt. This results in fewer particles trapped per launch attempt, but minimizes the number of wasted particles.
%\textcolor{red}{This sentence might come off a bit counterprodutive to the reader. We want to increase total capture rate not lower it.} 
For each launch, a single pulse was applied with a slew rate of 50 V/ms. The voltage (Vrms) was varied between 1 V and 3 V, and the pulse length between 1 ms and 300 ms, in order to consistently launch sufficient particles at each height. The driving frequency was kept between 140 and 160 kHz. %typically 150 kHz, which is one of the resonances of the piezo. 
The vacuum chamber pressure was maintained at $\sim$10 mbar. 
%\textcolor{red}{I would also include parameters like laser power and type of optical trap that was used.}

%Because we cannot rely on the motors and need to optically measure them, 
We determined the launch height using videos taken from an infrared camera positioned in the plane of the trap axis.
The image of the tip often appears saturated on the camera, making it appear artificially wide. 
%By measuring the unsaturated regions further up the capillary shaft and adding the pixel size uncertainty in quadrature, 
For this reason, we determined a positional uncertainty of $\pm$ 0.27 mm.

The trap was monitored after each launch attempt to record capture events. 
The total number of launches over the entire height range was $1850$. The resulting capture profile is shown in Fig.~\ref{fig:prob_graph}.
%To safely bound the data outside the $\pm$ 0.27 mm error limit, we sorted the launch heights into 0.55 mm bins. 
We normalized the total capture events in each 0.55 mm bin to report the expected number of catches per 100 launch attempts for that height range. %($N = 1850$). 
From these results, we determine an operating range where we capture at least 5 particles between 0.55 mm and 2.75 mm, with a peak near 2 mm, which results in the highest number of trapping events. We also note that we are able to trap particles at all heights attempted, and although not studied in this particular case, we have observed trapping in some setups for distances as large as 1 cm.
%\textcolor{red}{Consider removing this final sentence since in the capture profile, you are no longer distinguishing between aggregates and singles.}

% --- END OF SCOTT'S SECTION --
\section{Summary \& Outlook}
%Alexey, Mark(broader impact,UHV cryogenic trapping, trapping near surfaces without contamination)
%Emphasize advancements of this technique, ways to improve/next steps, potential applications in broader fields.

In summary, we have demonstrated a compact piezoelectric micropipette launcher for localized in-vacuum delivery of nano and microparticles into optical traps. %The same simple launcher has been integrated into multiple optical trapping setups, including a single-beam trap, non-interfering dual beam trap, and standing-wave dual beam trap showcasing the versatility and ease of integration of the setup. Using the micropipette launcher we successfully trapped silica spheres of $170\text{ nm}$, $300\text{ nm}$, 3 $\mu\text{m}$ diameter, as well as 6 $\mu\text{m}\times$ 0.2 $\mu\text{m}$ $\beta$-NaYF hexagonal prisms and $\sim 100$ nm diameter high-purity nanodiamonds. 
A typical micropipette loading of some of these particles leads to hundreds of successful trapping events with as high as 93\% efficiency. Even a filling with as little as one hundred thousand $\beta$-NaYF hexagonal prisms in the pipette still leads to multiple successful trapping events. We anticipate that successful launching is possible even with significantly lower amounts of particles. 
\newline
\indent
The main advantages of this loading method are its localized particle flux, compact geometry, ability to work with very small amounts of material, as well as compatibility with UHV and cryogenic setups. In addition, the method provides a large reservoir for repeated launching without breaking vacuum which can be important for applications such as sensing or matter wave interferometry with levitated particles, including possible space-based implementations \cite{Kaltenbaek:2012,Kaltenbaek_2023}. The localized source can help reduce contamination of nearby surfaces, electrodes, and optics. The proposed geometry is attractive for experiments where conventional aerosol loading is incompatible with vacuum or cryogenic operation, and where particles must be delivered close to a small trapping volume, optical cavity, ion trap, Paul trap, or chip-based trap.
\newline
\indent
Beyond levitated optomechanics, the method could be used as a general, dry, localized particle delivery tool for materials science, nanotechnology, micro-fabrication, surface science, aerosol science or dust physics in planetary science. The method might find applications for sample preparation in mass spectroscopy or electron microscopy. With suitable particle preparation, the method could also be explored for dry delivery of fragile or scarce biological, pharmaceutical or bio-molecular microparticles.
\newline
\indent
This technology can be improved in future in several aspects. In particular: (1) increasing the resonance frequency, amplitude and thus acceleration by optimizing the load mass, piezo choice and potentially matching the pipette's eigen-frequencies to exploit additional resonant behavior; (2) improvement of preparation techniques for higher single-to-aggregate trapping ratio; (3) extension to other types of particles such as functionalized particles or biological samples and (4) testing in other types of traps including magnetic and Paul traps. 
%improvements: UHV trapping, acceleration increase, improved preparation techniques for higher percentage of singles, extension to more classes of particles, testing in other types including magentic traps, increasing the number of launcher per filling and decreasing the required amount of particles per a successful trapping event

\section{Acknowledgements}

\textcolor{black}{We acknowledge Gavin Morley, Oliver Williams, Soumen Mandal, and Karishma Gokani for the aquisition, milling, and imaging of the nanodiamonds presented in this work. NaYF material used in this work was provided by Lars Forberger and Peter Pauzauskie. We are grateful for helpful discussions with Eduardo Alejandro.} A.G. ~acknowledges support from the W.M.~Keck Foundation, from NSF grants PHY-2409472 and PHY-2111544, DARPA, the John Templeton Foundation, the Gordon and Betty Moore Foundation Grant GBMF12328, DOI 10.37807/GBMF12328, the Alfred P.~Sloan Foundation under Grant No.~G-2023-21130, and the Simons Foundation. Alexey Grinin acknowledges the Alexander von Humboldt Foundation for the Feodor von Lynen Postdoctoral fellowship.

\section{Data Availability Statement}

The raw data reported in this paper as well as detailed design drawings and parts lists of the launching apparatus are available to interested parties upon reasonable request. 

\section{Statement of Author Contributions}
A.G. and A.D. conceptualized the method and performed experimental investigations.  M.N., S.G., S.K., S.G.E., S.B. performed experimental investigations. K.B.G. assisted with methodology development and validation. Z.W. and G.W. assisted with instrumentation development for the optical trapping studies with NaYF particles. A.A.G. supervised the research. All authors contributed to the writing of the manuscript.
%In this Appendix, we provide detailed design drawings of the pipette based particle loading mechanism...

%\begin{figure}[h]
    %\centering
    %\includegraphics[width=0.5\textwidth]%{cad_render.pdf}
    %\caption{Design drawing of the pipette launcher assembly with dimensions in mm for machined pieces.}
    %\label{fig:Drawing}
    %\caption{\textcolor{blue}{}}
%\end{figure}

\bibliography{main}% Produces the bibliography via BibTeX.

\end{document}